\documentclass[runningheads]{llncs}
\usepackage{url}

\usepackage{graphicx}
\usepackage{float}
\usepackage{multirow}
\usepackage{biblatex}
\usepackage[T1]{fontenc}
\begin{document}
\title{Automated Identification of Failure Cases in Organ at Risk Segmentation Using Distance Metrics: A Study on CT Data}
\titlerunning{Distance-Based Candidate Identification in Organ at Risk Segmentation} 
%
\author{Amin Honarmandi Shandiz \and  Attila Rádics\and Rajesh Tamada \and Makk Árpád \and Karolina Glowacka \and Lehel Ferenczi \and Sandeep Dutta \and Michael Fanariotis}
\authorrunning{A. Honarmandi Shandiz et al.}
%
\institute{GE HealthCare\\
\email{\{Amin.Honarmandishandiz, Attila.Radics, Rajesh.Tamada, Arpad.Makk1, Karolina.Glowacka1, Sandeep.Dutta, Michail.Fanariotis   \}@ge.com, Lehel.Ferenczi@med.ge.com}
}
\maketitle              
\begin{abstract}
Automated organ at risk (OAR) segmentation is crucial for radiation therapy planning in CT scans, but the generated contours by automated models can be inaccurate, potentially leading to treatment planning issues. The reasons for these inaccuracies could be varied, such as unclear organ boundaries or inaccurate ground truth due to annotation errors. To improve the model's performance, it is necessary to identify these failure cases during the training process and to correct them with some potential post-processing techniques. However, this process can be time-consuming, as traditionally it requires manual inspection of the predicted output. This paper proposes a method to automatically identify failure cases by setting a threshold for the combination of Dice and Hausdorff distances. This approach reduces the time-consuming task of visually inspecting predicted outputs, allowing for faster identification of failure case candidates. The method was evaluated on 20 cases of six different organs in CT images from clinical expert curated datasets. By setting the thresholds for the Dice and Hausdorff distances, the study was able to differentiate between various states of failure cases and evaluate over 12 cases visually. This thresholding approach could be extended to other organs, leading to faster identification of failure cases and thereby improving the quality of radiation therapy planning.

\keywords{Anomaly Detection, Quality Assurance, Organ at risk segmentation, Failure case identification, Distance metrics, Model performance, Radiation therapy planning, Thresholding }
\end{abstract}
\section{Introduction}
Organ at risk (OAR) segmentation is an essential step in radiation therapy planning. Accurate delineation of OARs can help to spare these critical structures from high doses of radiation and prevent unwanted side effects~\cite{van2012functional}. The traditional method for OAR contouring is manual contouring, which is a time-consuming and labor-intensive process. Manual contouring can also be subjective and prone to inter-observer variability, which can affect the accuracy and consistency of the contouring~\cite{brouwer20123d,brouwer2015ct,breunig2012system}.
To address these issues, automatic contouring methods have been developed in recent years. These methods use algorithms and machine learning techniques to automatically segment OARs from medical images. Recently, deep learning methods such as Convolutional neural networks have played a significant role in various aspects specially when the data is images~\cite{honarmandi2021voice,shandiz2021improving,shandiz2021neural,csapo2022optimizing,toth20203d,yu2021reconstructing}.

According to a study conducted by Fugan et al~\cite{fung2020automatic} an auto-contouring (AC) system was evaluated and found to be in agreement with manual contours within the inter-observer uncertainty level. Additionally, Altman et al~\cite{altman2015framework} observed that there were no significant dosimetric differences between treatment plans using AC software versus manually contouring. These findings suggest that despite not always being perfect, auto-contours can still be considered clinically acceptable.
However, the performance of automatic contouring methods can vary depending on the imaging modality, image quality, and complexity of the OARs being segmented~\cite{fung2020automatic}. As such, it is important to evaluate the accuracy and robustness of automatic contouring methods before their implementation in clinical practice.

A conventional approach to enhance the performance of OAR models is to improve the accuracy of the model's predictions. To achieve this, it is necessary to identify cases where the model fails to produce accurate predictions. In the case of OAR models, this requires a visual inspection of the predicted contours to determine if they are closely overlapping the ground truth. Cases where there is unsatisfactory overlap between the predicted and ground truth contours, or where anomalies exist in the predicted images, are selected as failure cases. These cases are then subjected to further analysis to identify the cause of the failure, which involves consultation with experts. However, this approach can be time-consuming, especially when there are many candidate cases to test and no predefined set of candidates to prioritize for failure analysis.

In order to address this issue, our paper proposes a method that utilizes distance metrics to automatically identify candidate cases for failure analysis. By analyzing 20 cases for 6 organs, we determined a threshold that could effectively distinguish these candidates. Instead of manually checking the predicted contours against the ground truth, this approach could save time by automatically identifying potential failure cases based on the HD and Dice metrics~\cite{dice1945measures}. We found that different organs had different values for the proposed threshold, and this method could help decrease the time needed to find failure cases.

\section{Methods and Materials}
\begin{figure}[ht]
\centering
\includegraphics[width=0.8\textwidth]{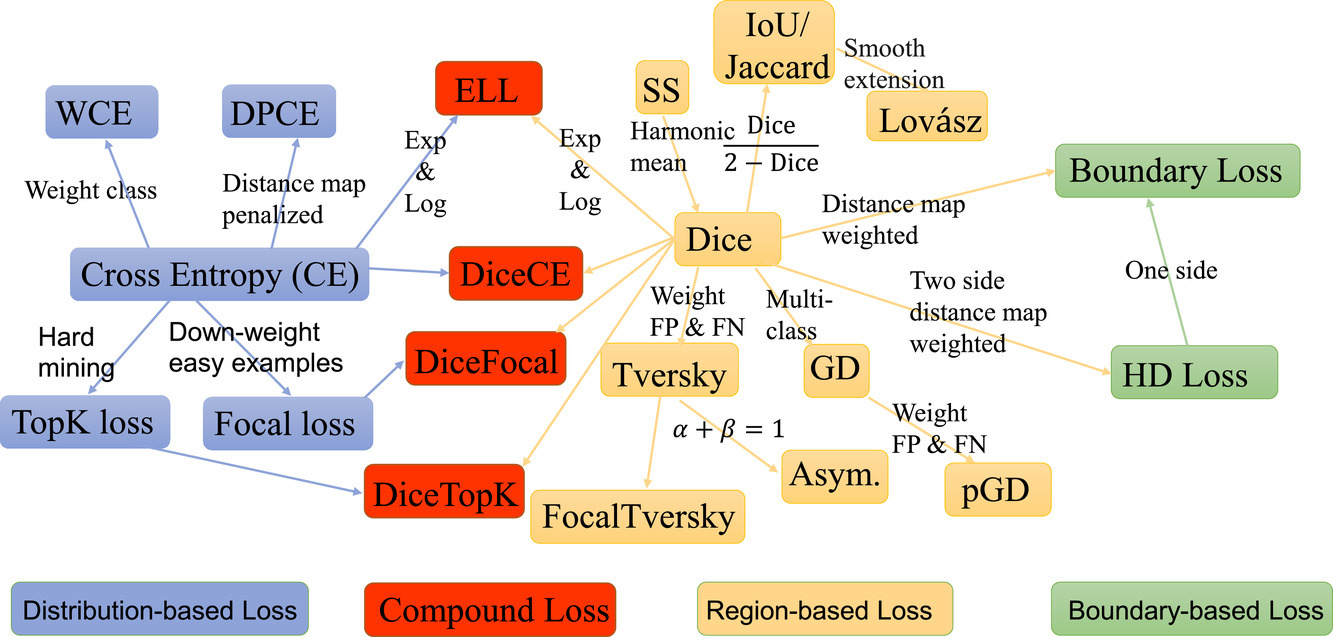}
\caption{The Figure shows loss functions for typical 3D segmentation tasks in four different category ~\cite{he2020multi}.} \label{fig:loss}
\end{figure}

Loss functions play an important role in various tasks. In image segmentation tasks, the role of the loss function is distinguished by measuring the difference between the predicted and ground truth segmentation maps. They guide the training process to optimize model performance. Some common loss functions in common 3D segmentation tasks can be seen in figure~\ref{fig:loss} by Jun Ma et al~\cite{ma2021loss}. 
According to the study conducted by Shruti et al. Common loss functions in image segmentation are different and depend on various parameters, although the popular used ones can be cross-entropy loss, dice loss, Hausdorff distance (HD), etc.~\cite{jadon2020survey}.
In the context of OAR segmentation tasks, the choice of loss functions depends on task-specific requirements such as the nature of the data, the desired characteristics of the segmentation output, and the specific task.

\subsection{Dice and Hausdorff Distance Metrics}
To evaluate the accuracy of the OAR segmentation, we used the Dice and Hausdorff distance metrics. The Dice coefficient measures the similarity between two sets of binary data, while the Hausdorff distance measures the maximum distance between two sets of points. We computed the Dice and Hausdorff distance between the predicted segmentation masks and ground truth masks for each of the six OARs.

The Dice distance, also known as the Dice coefficient, is a widely used metric for evaluating the similarity or overlap between two sets. In the context of medical image analysis, it is often used to assess the accuracy of segmentation algorithms that delineate the boundaries of anatomical structures or regions of interest.

The Dice distance is calculated as the ratio of the intersection of two sets to their combined size, and ranges from 0 to 1, where a value of 1 indicates perfect overlap and a value of 0 indicates no overlap. 
The formula for Dice distance is:

\begin{equation}
    Dice(A,B) = \frac{2\left| A \cap B \right|}{\left| A \right| + \left| B \right|}
\end{equation}\label{eguit:dice}
where $\cap$ represents the intersection of the predicted and actual volumes.
The Hausdorff distance is another metric used to evaluate the accuracy of segmentation algorithms. It measures the maximum distance between any point on the surface of one volume and the closest point on the surface of the other volume. In other words, it quantifies the "worst-case scenario" of disagreement between two sets.
\noindent The formula for Hausdorff distance is:
\begin{equation}
    Hausdorff(A,B) = \max\{h(a,B),h(b,A)\}
\end{equation}\label{equit:hd}

where a and b are points on the surfaces of A and B, respectively, and h(a,B) and h(b,A) are the distances from a to the closest point on B, and from b to the closest point on A, respectively.

The Hausdorff distance is often used in conjunction with the Dice distance to provide a more complete picture of the accuracy of a segmentation algorithm. While the Dice distance measures the degree of overlap between two sets, the Hausdorff distance provides information about the extent and location of any discrepancies between the two sets.

\subsection{Dataset}

We have used two datasets, -i): one containing five organs-at-risk (OARs) that are commonly segmented in radiation therapy planning from an internal dataset and -ii) another one contains one organ from a public dataset for lymph node~\cite{seff20142d,roth2014new,seff2015leveraging}. The internal dataset (acquired by our clinical partners) consists of 20 CT images depicting different areas, scanned using different CT protocols, with slice thickness between 0.5 and 3 mm. These scans were acquired on volunteers, using various multi-slice CT scanners such as (Discovery 750HD, Optima540, etc). These scans, unlike the ones in the public dataset, are not uniform, and differ in CT image parameter settings (e.g. resolution, pixel spacing). 

\subsection{Model Architecture}

\begin{figure}[ht]
\centering
\includegraphics[width=1.0\textwidth]{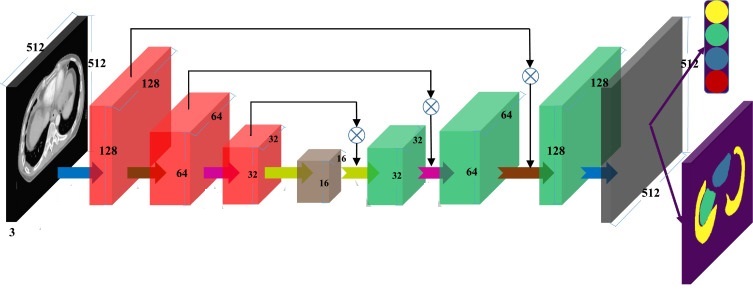}
\caption{The figure shows the 3D auto-encoder model for organ at risk segmentation~\cite{he2020multi}.} \label{fig:model}
\end{figure}

The figure~\ref{fig:model} shows a schematic diagram of an auto-encoder model, which is a type of neural network used for tasks such as dimensional reduction and feature extraction. The model consists of two parts: the encoder and the decoder. The encoder takes the input data and compresses it into a lower-dimensional representation or latent space, while the decoder takes this representation and reconstructs the original data. This model can be used for various applications, including image and signal processing.

In our paper, organ contours were created using a novel DL auto segmentation model (DLAS v1.0, GE HealthCare), that the auto encoder structure is like figure~\ref{fig:model} used in~\cite{he2020multi}. We employed the encoder part of the model as a feature extractor to obtain a compressed representation of the input images that captures the salient features related to the organs of interest. This compressed representation was then fed into a decoder to generate the organ segmentation contours. This approach allowed us to improve the accuracy and efficiency of the segmentation task compared to manual contouring or other traditional segmentation methods.

\subsection{Experimental Setup}\label{sec_class}
Firstly, the contours were produced and reviewed by experts radiologist for five different organs including breasts, Chiasma, Brainstem, Trachea and Femurs. for each organ there are 20 CTs with their ground truth. Then, we converted all of the data to nrrd format and fed them to the 3D auto-encoder model like figure~\ref{fig:model} with train, evaluation and test partitions respectively 10,5,5 which were chosen for the model and batch size of 5.
\begin{figure}[ht]
\centering
\includegraphics[width=1.0\textwidth]{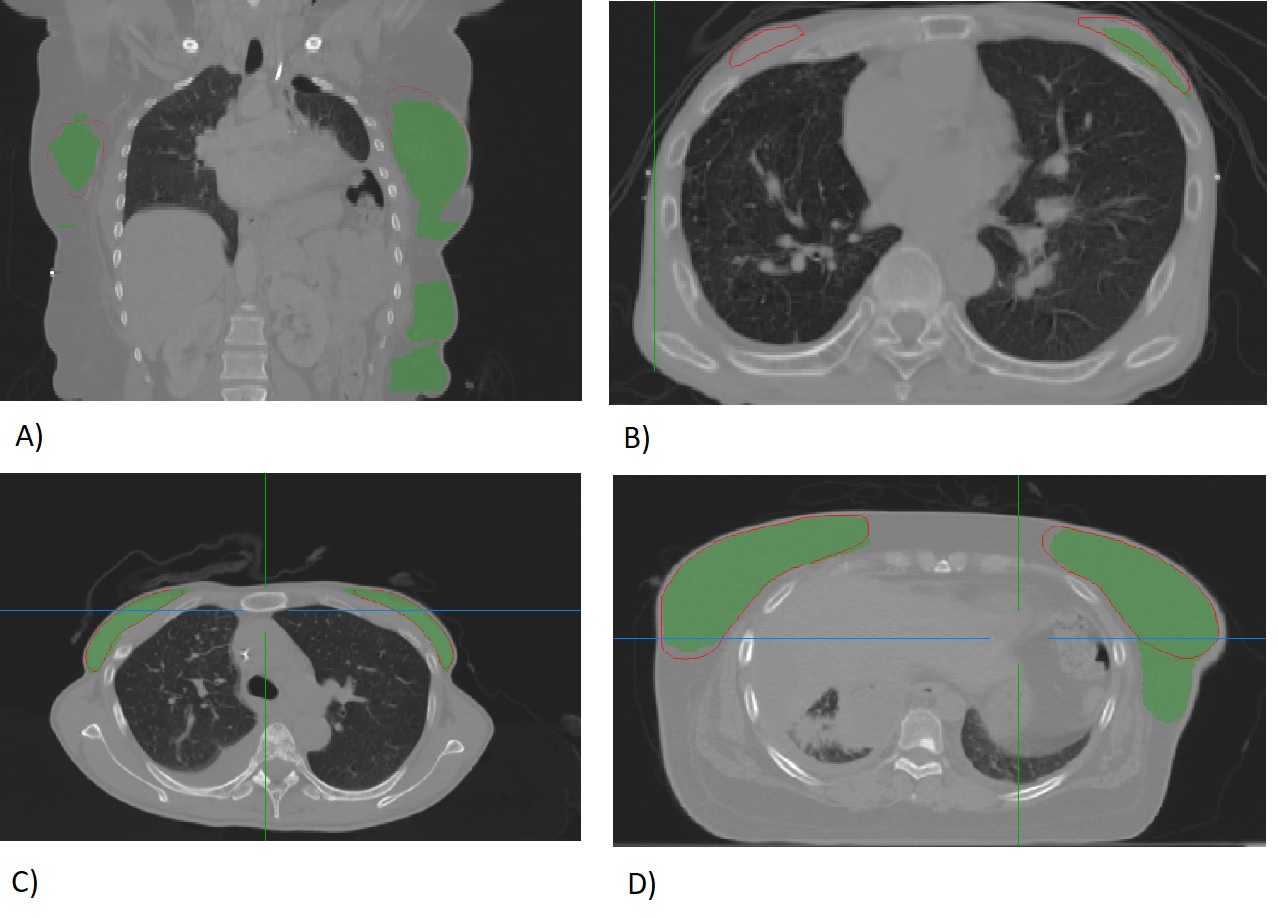}
\caption{The figure summarizes four situations expressed in Table~\ref{anomaly_detection} for Breasts organ segmentation. The four situations include: A) Shows over segmentation of the model, extra predictions exists in the output beside the region of interest,  B)Extra predictions with less overlap with ground truth for the left breast and under segmentation for the right breast organ in some slices, C) good predictions, D) model has good predictions but there are extra predictions(over segmentation) in the left breast. Ground Truth contours are distinguished by red line and predictions are distinguished by green.} \label{fig:breast}
\end{figure}\
For training the model we used Dice as loss function. After contours were predicted by the model then two metrics including Hausdorff and Dice~\cite{maiseli2021hausdorff,karimi2019reducing} used to evaluate the model prediction. 
After visualizing the predicted contours, we compared the results with ground truth to find if there are failure prediction exist in the predicted values. Afterward, we visualize the predicted failed cases versus good cases as presented in figure~\ref{fig:breast}.

As it is showed in figure~\ref{fig:breast} for breast organ there are situations where the predicted organs does not overlap with the ground truth or there might be extra lesion predicted outside of the expected organ region. To find these cases we use the characteristics of image distance measures such as Dice and Hausdorff distances. As the Dice's characteristic~\cite{dice1945measures} is suitable to find the overlapped contours and base on characteristics of Hausdorff distance~\cite{huttenlocher1993comparing} that is considered for distance calculation, we use this metric to find the regions which are predicted far from the ground truth as it is sensitive to the outliers.

\begin{figure}[ht]
\centering
\includegraphics[width=1.0\textwidth]{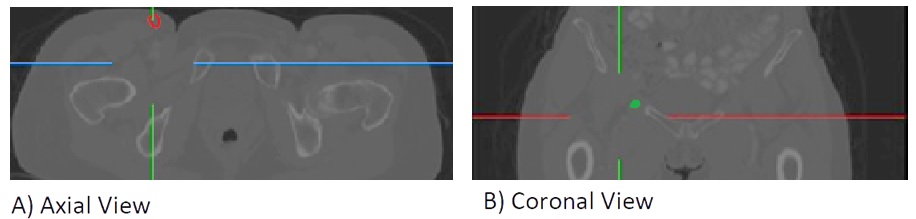}
\caption{The figure shows the results of lymph node contouring in CT images using a deep learning model.  } \label{fig:lymphnode}
\end{figure}
 
We also analyised lymph node data(see fig~\ref{fig:lymphnode}) containing abdominal and mediastinal lymph nodes which are publicly available total of 388 mediastinal lymph nodes in CT images of 90 patients and a total of 595 abdominal lymph nodes in 86 patients (for more info see~\cite{seff20142d,seff2015leveraging,roth2014new}).



\section{Results and Discussion}\label{res}
After calculating Dice and Hausdorff distances for predicted versus ground truth, the purpose was to find the less predicted contours in some slices by low Dice score values. Also, analyzing the HD values could determine if the output is over segmented or not. As you can see in table~\ref{anomaly_detection} with different combination of Dice and Hausdorff values we are able to decide if output contours need to consider for post-processing.

We found that setting a threshold for Dice and Hausdorff distance produced for these six organs could help to distinguish the failure cases, identifying a significant number of candidate cases while keeping the false positive rate low. We also found that in most cases, the identified failures were due to under segmentation between the predicted and ground truth contours, or due to over segmentation by the model.

We compared our approach to a baseline method that relied on visual inspection alone and found that our method could identify candidate cases with higher speed as it is not needed to visually check them. Our approach is also could be useful in other organs by finding their threshold values.
\begin{table}[ht]
\centering
\caption{The table summarizes the different situations for anomaly detection in medical imaging based on the Hausdorff distance and Dice similarity coefficient for Breast. The table provides an explanation for each situation based on the predicted results.}\label{anomaly_detection}

  \begin{tabular}{|p{0.4\linewidth}|p{0.5\linewidth}|}
    \hline
    
    Situation & Explanation \\
    \hline
    \hline
    HD $<$ 6 and Dice $\sim$ 0.9 & Good prediction \\
    \hline
    6 $<$ {HD} $<$ 100 and Dice $\sim$ 0.9 & Good overlap prediction but there are extra predictions connected and close to GT\\
    \hline
    Hd $>$ 100 and Dice $\sim$ 0.9 & Good prediction but there are extra predictions which are far from respective organ \\
    \hline
    Hd $>$ 6 and Dice $<$ 0.8 & There are extra predictions in slices or far from organ and there is led overlap for some ground truth contours \\
    \hline
    Hd $<$ 6 and Dice $<$ 0.8 & There is no extra pridicted contours or lesion but no predicted contours for some GT in some slices \\
    \hline
  \end{tabular}

\end{table}
In summary, our research highlights the usefulness of employing a range of distance metrics for detecting inferior or unsatisfactory segmentation predictions in organ-at-risk segmentation, which can enhance the performance of models and facilitate model failure analysis in medical imaging. It should be noted that the appropriate threshold would vary for different organs, and in respect of the size of the organs and contrast and other factors it could change.

In order to evaluate the performance of our proposed approach, First we used the Table~\ref{anomaly_detection} values to identify good cases with good overlap and less out of box prediction. Then, we select randomly 12 cases to visually inspect the generated contours. In this step we could see that out of 12 cases which were randomly selected by this approach, the selected ones have less anomaly inside the predicted contours.

As you can see in Table~\ref{anomaly_detection} to evaluate the model's performance to detect the anomalies in the predicted contours, we have considered different situations based on the combination values of Hausdorff Distance and Dice coefficient. When Hausdorff distance is less than 6mm and the Dice coefficient is around 0.9, we consider the prediction to be a good one. In the case of Hausdorff between 6 and 100 and the Dice coefficient around 0.9, we have identified the prediction as having a good overlap with the ground truth contours but also having over segmentation because of high value in HD. For Hausdorff distance greater than 6 and the Dice coefficient around 0.9, we acknowledge the prediction as a good one, but with extra predictions located far from the respective organ. Moreover, when Hausdorff is greater than 6 and the Dice coefficient is less than 0.8, we determine that there are over segmentation in slices or far from the organ and there is lack of overlap between ground truth and prediction in some of the slices. Finally, when HD is less than 6 and Dice also is less than 0.8 percent then there are less predicted values for GT and less over segmentation because the HD value is small.

\begin{table}[ht]
\centering
\caption{Thresholds for Hausdorff distance and Dice score for organ segmentation. The table lists the recommended thresholds for breast and lymph node segmentation based on Hausdorff distance and Dice score metrics. The values are based on expert review of a sample dataset.}\label{threshhold}
\begin{tabular}{|l|c|c|}
\hline
Organ & Dice score & Hausdorff Distance \\
\hline
Breast & 0.9 & 6 mm \\
\hline
Lymph Node & 0.95 & 0.5 mm \\
\hline
Femur & 0.9 & 11 mm \\
\hline
Trachea & 0.85 & 19 mm \\
\hline
Chiasma & 0.66 & 3.4 mm \\
\hline
Brainstem & 0.8 & 10 mm \\
\hline
\end{tabular}

\end{table}

\begin{figure}[ht]
\centering
\includegraphics[width=1.0\textwidth]{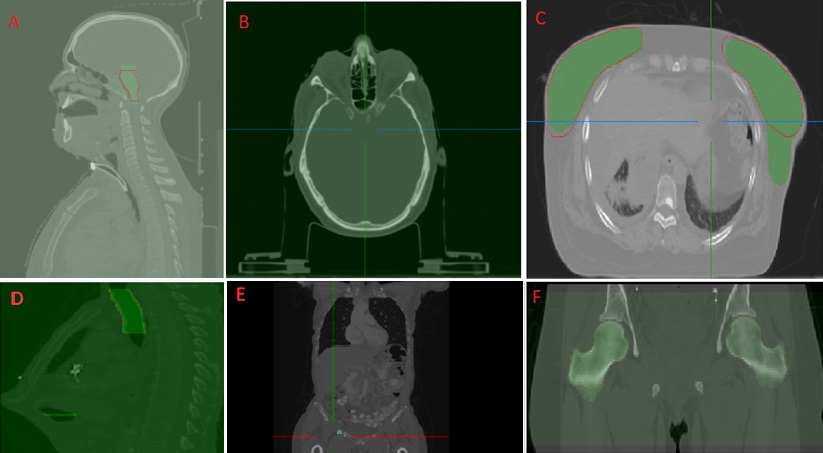}
\caption{
The figure shows six contoured organs in different boxes, including brainstem, chiasma, breast, trachea, lymph node and femur. These organs are important in radiation therapy planning as they are at risk of damage during treatment. Accurate segmentation of these organs is essential for successful treatment, and auto-contouring models are increasingly being used for this purpose.  } \label{fig:organs}
\end{figure}
 As it is showed in fig~\ref{fig:organs} we visualized different organs and showing their inaccurate prediction that needs to find automatically by our method.
We analyzed 20 CTs of 6 organs and we could find threshold values for Dice and HD for these organs as you can see in Table~\ref{threshhold} for respective organs. Then we compared 12 cases of each organs which chosen randomly by setting this threshold and compare them visually to find out if this method could distinguish the candidate failure cases.

\section{Conclusion}\label{conc}
In conclusion, our paper proposed a method for identifying potential failure cases in organ at risk segmentation using distance metrics, which is a combination of Dice and Hausdorff distances. By analyzing 20 cases of six different organs and visually comparing the ground truth and predicted contours, we were able to determine a threshold for each organ that could potentially decrease the time required for identifying failure cases. Our findings suggest that this approach could be useful for improving the performance of auto contouring models in radiation therapy planning in order to speed up selecting failure cases for post-processing. In future research, we plan to explore the use of this method for anomaly detection after model training and for improving the model performance during clinical decision-making.
\section{Acknowledgments}
We would like to thank all the contributors from the Deep Learning Auto Segmentation  (DLAS) project for facilitating this study. 

\printbibliography
\end{document}